\documentclass[pre,twocolumn,10pt,nofootinbib,superscriptaddress,showpacs]{revtex4}

\usepackage{graphicx}
\usepackage{amsmath}
\usepackage{comment}
\usepackage{times}
\usepackage{multirow}
\usepackage[usenames]{color}
\usepackage{chemarr}
\usepackage{ifthen}
\newcommand{\SSA}[1]{%
    \ifthenelse%
        {\equal{#1}{l}}%
            {stochastic simulation algorithm}
            {SSA}
}
\newcommand{\ES}[1]{%
    \ifthenelse%
        {\equal{#1}{l}}%
            {exact-stochastic}%
            {ES}%
}
\newcommand{\DM}[1]{%
    \ifthenelse%
        {\equal{#1}{l}}%
            {direct method}%
            {DM}%
}
\newcommand{\FRM}[1]{%
    \ifthenelse%
        {\equal{#1}{l}}%
            {first-reaction method}%
            {FRM}%
}
\newcommand{\NRM}[1]{%
    \ifthenelse%
        {\equal{#1}{l}}%
            {next-reaction method}%
            {NRM}%
}
\newcommand{\NSM}[1]{%
    \ifthenelse%
        {\equal{#1}{l}}%
            {next-subvolume method}%
            {NSM}%
}
\newcommand{\PLA}[1]{%
    \ifthenelse
        {\equal{#1}{l}}
            {partitioned-leaping algorithm}
            {PLA}%
}
\newcommand{\SPLA}[1]{%
    \ifthenelse
        {\equal{#1}{l}}%
            {spatial partitioned-leaping algorithm}%
                {SPLA}{}%
}
\newcommand{\SB}[1]{%
    \ifthenelse%
        {\equal{#1}{l}}%
            {species-based}%
            {SB}%
}
\newcommand{\RB}[1]{%
    \ifthenelse%
        {\equal{#1}{l}}%
            {reaction-based}%
            {RB}%
}
\newcommand{\MLB}[1]{%
    \ifthenelse{\equal{#1}{s}}%
        {MLB}%
        {\ifthenelse{\equal{#1}{fauthor}}%
            {Marquez-Lago}%
            {\ifthenelse{\equal{#1}{full}}%
                {Marquez-Lago and Burrage}{}%
            }%
        }%
}
\newcommand{\RBK}[1]{%
    \ifthenelse{\equal{#1}{s}}%
        {RBK}%
        {\ifthenelse{\equal{#1}{fauthor}}%
            {Rossinelli}%
            {\ifthenelse{\equal{#1}{full}}%
                {Rossinelli~et~al.}{}%
            }%
        }%
}
\begin{document}

\title{Accurate implementation of leaping in space: The spatial partitioned-leaping algorithm}

\author{Krishna A. Iyengar}
\email{kai8@cornell.edu}%
\affiliation{School of Theoretical \& Applied Mechanics, Cornell University, Ithaca, NY 14853, USA}%

\author{Leonard A. Harris}
\altaffiliation[Current address: ]{Department of Computational Biology, University of Pittsburgh School of Medicine, Pittsburgh, PA 15260, USA.}
\email{lharris@pitt.edu}%
\affiliation{School of Chemical and Biomolecular Engineering, Cornell University, Ithaca, NY 14853, USA}

\author{Paulette Clancy}%
\email{pqc1@cornell.edu}%
\affiliation{School of Chemical and Biomolecular Engineering, Cornell University, Ithaca, NY 14853, USA} %

\date{\today}

\begin{abstract}
There is a great need for accurate and efficient computational approaches that can account for both the discrete and stochastic nature of chemical interactions as well as spatial inhomogeneities and diffusion.  This is particularly true in biology and nanoscale materials science, where the common assumptions of deterministic dynamics and well-mixed reaction volumes often break down. In this article, we present a spatial version of the \PLA{l} (\PLA{s}), a multiscale accelerated-stochastic simulation approach built upon the $\tau$\/-leaping framework of Gillespie. We pay special attention to the details of the implementation, particularly as it pertains to the time step calculation procedure. We point out conceptual errors that have been made in this regard in prior implementations of spatial $\tau$\/-leaping and illustrate the manifestation of these errors through practical examples. Finally, we discuss the fundamental difficulties associated with incorporating efficient exact-stochastic techniques, such as the \NSM{l}, into a spatial-leaping framework and suggest possible solutions.
\end{abstract}

\maketitle

\section{Introduction} \label{sec:introduction}

In our attempts to understand the behavior of systems at increasingly small scales, the importance of random fluctuations, or noise, is becoming increasingly apparent. Indeed, the phenomenon is the subject of great interest in a variety of diverse fields, including cellular biology \cite{Arkin98, McAdams99, Elow02, Fedo02, Rao02, Longo06, Samo06:STKE}, semiconductor processing \cite{Plumm95, Roy05} and heterogeneous catalysis \cite{Johanek04}.

From a computational perspective, incorporating the effects of stochasticity into models of physical processes requires moving beyond traditional continuum-deterministic approaches, such as ordinary differential equations (ODEs), and using one of a variety of stochastic methods. Within the purview of chemical kinetics, a popular technique is Gillespie's \emph{\SSA{l}} (\SSA{s}) \cite{Gillesp76, Gillesp77, Gillesp07}. The method is extremely accurate, easy to implement and has found widespread use in computational systems biology.  Its downside however, is speed, and the algorithm can become prohibitively slow due to its one-reaction-at-a-time nature \cite{Endy01, Gillesp01}.

This fact has spawned considerable effort, from a variety of directions, to develop methods for overcoming this inherent limitation of \ES{l} approaches. A particularly popular type of accelerated-stochastic approach is ``$\tau$\/-leaping", originally devised by Gillespie \cite{Gillesp01} and expanded upon by numerous investigators \cite{Gillesp03, Rath03, Cao05:negPop, Cao06:newStep, Cao07, Tian04, Chatt05, Auger06, Cai07, Petti07, Peng07, Rath07, Ander08, Xu08, Leier08}, including ourselves \cite{Harris06}.  In general, leaping methods have proved quite successful in overcoming some, but not all, of the problems plaguing exact-stochastic simulation methods \cite{Gillesp07}.

All of the methods cited above operate under the assumption that the volume within which the reactions are ``firing" is ``well mixed." In more precise terms, the assumption is that the time scale of diffusion is fast enough so that all entities (e.g., molecules) of the same species have equal probability of reacting at any given point in time. However, it is not hard to imagine situations where this assumption breaks down. In solid-state systems, for example, diffusion is much slower than in fluids and the local environment seen by a dopant atom, say, plays a much larger role in its dynamics \cite{Roy05}. In biology, both eukaryotic \cite{Mulcahy08} and prokaryotic \cite{Callaway08} cells have intricate internal structures that act to localize certain interactions and processes. The sheer size of cellular components also leads to a highly crowded and definitively \emph{non-well-mixed\/} intracellular environment \cite{Ellis01, Zhou08}.

In situations such as these, methods that account for spatial inhomogeneity and diffusion are needed.  In the extreme case, it may be necessary to track the fates of individual entities, or ``agents" \cite{Shimizu03, Rodrig06}.  However, a more common situation is one where the system of interest can be partitioned into multiple smaller domains, or ``subvolumes."  Each subvolume is assumed to be well-mixed and coupled to neighboring subvolumes via a jump-diffusion processes.  Various extensions of the SSA have been successfully implemented along these lines \cite{Fricke95, Grach01, Elf04, Hattne05, Bernst05}.  General overviews of both agent- and subvolume-based spatial-stochastic simulation approaches applied in biology and materials science can be found in Refs.~\cite{Takaha05, Lemerle05, Andrews06, Dobrzy07, Chatt07}.

In spatially inhomogeneous systems, the shortcomings of the \ES{l} approach are intensified. In general, each subvolume is given local copies of each reaction and diffusion event.  Thus, the number of possible events in the system increases significantly with increasing number of subvolumes, often making SSA-like methods infeasible. A partial solution to this problem lies with the leaping methods.  While the number of events in the system remains unchanged (and hence still a potential problem), spatial leaping methods achieve accelerations by allowing all reaction and diffusion events to fire multiple times at each simulation step. We are aware of two implementations of leaping algorithms along these lines, those of \MLB{full}\ \cite{Lago07} and \RBK{full}~\cite{Rossi08}. \MLB{full} propose a method that is a leaping analogue to the well-known ``\NSM{l}" (\NSM{s}) \cite{Elf04, Hattne05}, an efficient spatial \SSA{s} variant. \RBK{full} present a more straightforward extension of leaping in space that considers reaction and diffusion events separately.

In this article, we present a spatial implementation of our own method, the \emph{\PLA{l}} (\PLA{s}) \cite{Harris06}. Our implementation is similar in spirit to the methods of \MLB{fauthor}, and of \RBK{fauthor}, but differs in some important ways.  In particular, we take special care with regards to the calculation of time steps. We point out some conceptual errors that were made in this regard in refs.~\cite{Lago07} and \cite{Rossi08} and demonstrate, through numerical examples, how these errors may affect accuracy and efficiency.  We show that, in some cases, the \SPLA{l} (\SPLA{s}) is faster than these methods and at least as accurate. In other cases, SPLA is slower but significantly more accurate. In yet other cases there is little difference. We explain the origins of this differential behavior and its consequences for practical applications of the methods. Finally, we discuss the fundamental difficulties associated with incorporating exact-stochastic approaches like the \NSM{s} into a spatial-leaping framework and suggest possible strategies for overcoming them.

In Sec.~\ref{sec:background}, we present an overview of relevant exact- and accelerated-stochastic simulation methods for both homogeneous (well-mixed) and inhomogeneous systems that set the stage for the new \SPLA{s} approach in Sec.~\ref{sec:SPLA}.  Sec.~\ref{sec:examples} shows results from three simple example systems that exemplify the gains in accuracy and efficiency achieved by the method.  Finally, we conclude in Sec.~\ref{sec:discussion} with a discussion of these results and their implications for future extensions of leaping methods.

\section{Background} \label{sec:background}

We consider a chemically reactive system of fixed volume $\Omega_V$ and constant temperature that is partitioned into $L$\/ well-mixed subvolumes $\mathbf{V}\!=\!\{V_1,\ldots,V_L\}$.  Each subvolume $V_l$\/ has a fixed volume $\omega_l$\/ ($\sum_{l^\prime=1}^L \omega_{l^\prime}\!=\!\Omega_V$) and is adjacent to $\Gamma_l$\/ ($\leq\!L\!-\!1$) neighboring subvolumes $\mathbf{C}_l\!=\!\{C_{l1},\ldots,C_{l\Gamma_l}\}$. In principle, each $V_l$\/ contains a \emph{unique\/} set of $N_l$\/ molecular species $\mathbf{S}_l\!=\!\{S_{l1},\ldots,S_{lN_l}\}$ that participate in $M_l$\/ unique reactions $\mathbf{R}_l\!=\!\{R_{l1},\ldots,R_{lM_l}\}$. We assume that all $N_l$\/ species can diffuse into and out of all $\Gamma_l$\/ neighboring subvolumes. Thus, each $V_l$\/ has $N_l\Gamma_l$\/ \emph{outgoing\/} diffusion events $\mathbf{D}_l\!=\!\{D_{l1},\ldots,D_{lN_l\Gamma_l}\}$ associated with it as well as $N_l\Gamma_l$\/ \emph{incoming\/} diffusion events $\mathbf{\widetilde{D}}_l\!=\!\{\widetilde{D}_{l1},\ldots,\widetilde{D}_{lN_l\Gamma_l}\}$. It is important to recognize that each $\widetilde{D}_{l\mu}$\/ is a \emph{reference\/} to an outgoing diffusion event from an adjacent subvolume. All together, there are a total of $M_l\!+\!2N_l\Gamma_l$\/ reaction and diffusion events associated with each $V_l$\/.  We thus define, without loss of generality, the event vector $\mathbf{E}_l\!=\!\mathbf{R}_l\!+\!\mathbf{D}_l\!+\!\mathbf{\widetilde{D}}_l$.

The state of the system is represented by the vector $\mathbf{X}(t)\!=\!\sum_{l^\prime=1}^L \mathbf{X}_{l^\prime}(t)$, where $X_{li}(t)$ is the population of species $S_i$\/ in subvolume $V_l$\/ at time $t$\/, $i\!\in\!\{1,\ldots,N_l\}$. Each event channel $E_{l\mu}$\/ is associated with a propensity function $a_{l\mu}(\mathbf{X}(t))$ (the stochastic analogue to the deterministic reaction rate) and a stoichiometry vector $\mathbf{z}_{l\mu}\!=\!\{z_{l\mu 1},\ldots,z_{l\mu N_l}\}$, $\mu\!\in\!\{1,\ldots,M_l\!+\!2N_l\Gamma_l\}$ (see \cite{note:indices}).

\subsection{Exact-stochastic methods} \label{sec:exactStoch}

\subsubsection{Well-mixed systems} \label{sec:wellMixed}

Gillespie's  \SSA{s} operates within a fully well-mixed system (i.e., $L\!=\!1$) \cite{Gillesp76, Gillesp77}. The approach determines \emph{when\/} the next reaction will fire in the system and of \emph{which\/} type it will be. Two mathematically equivalent approaches were presented for accomplishing this: the \DM{l} (see \cite{Gillesp07} for details) and the \FRM{l}. The \FRM{l} determines when each reaction in the system would fire \emph{if it were the only reaction present in the system\/} and then chooses $\tau$\/ as the smallest of these values and $\mu$\/ as the corresponding reaction. Such ``tentative" next-reaction times are calculated via
\begin{equation}
    \tau_\mu^\mathrm{exact} = -\ln(r_\mu)/a_\mu(t),
    \label{eq:FRM-tauMu}
\end{equation}
where $r_\mu$\/ is a unit-uniform random number. As originally formulated, the \FRM{l} requires $M$\/ unit-uniform random number generations at each simulation step, $M\!-\!1$ of which are discarded before proceeding on to the next step. An improvement upon this approach is Gibson and Bruck's \NRM{l} (\NRM{s}) \cite{Gibson00}. The \NRM{l} basically uses a rigorous random-variable transformation formula to reuse the generated random numbers in the next time step. This reduces the number of random number generations per time step to exactly one, along with  $M^\prime\!-\!1$ calculations of
\begin{equation}
    \tau_\mu^\mathrm{exact} = \left(a_\mu^\prime(t) / a_\mu(t)\right) (\tau_\mu^{\mathrm{exact}^\prime} - \tau^\prime),
    \label{eq:NRM-tauMu}
\end{equation}
where the unprimed and primed quantities signify new and old values, respectively.

\subsubsection{Inhomogeneous systems} \label{sec:inhomogeneous}

The \DM{l} and \FRM{l} essentially constitute two ends of a spectrum with regards to the grouping of reactions. In the \DM{l}, the entire system of reactions is basically considered to be one large group. In the \FRM{l} (and \NRM{s} by extension), each reaction is considered individually, i.e., as a group of one. Thus any method intermediate between these two is also a theoretically sound approach \cite{Gillesp07}. From a practical point of view, this means we are free to group reactions into subgroups as we see fit.  We can then choose among those subgroups using the \DM{l} or \FRM{l} (or \NRM{s} or any other equivalent method, e.g., \cite{Slepoy08, Schulze08}) and then choose within the subgroup in the same way.  Moreover, we can nest the subgroups into as many levels as we like if we find it convenient to do so.

Such a procedure has the effect of parsing out the computational load into multiple stages and can, in many cases, significantly improve the efficiency of the method. A well-known such approach is Elf and Ehrenberg's \NSM{l} (\NSM{s}) \cite{Elf04, Hattne05}, a spatial \SSA{s} variant that discretizes space into subvolumes and groups events (reaction and diffusion) based on the subvolume within which they reside.  The \NSM{s} operates by calculating the \emph{summed\/} propensities $a_{l0}(t)\!\equiv\!\sum_{\nu=1}^{M_l+N_l\Gamma_l} a_{l\nu}(t)$ for all subvolumes $l\!\in\!\{1,\ldots,L\}$. The subvolume within which the next reaction will fire is then identified using a heap search as in the \NRM{s} \cite{Gibson00} and the identity of the firing reaction within the subvolume using a linear search as in the \DM{l} \cite{Gillesp76}.  This two-level approach significantly reduces the computational effort relative to a straightforward heap or linear search over all events in the domain.

\subsection{Leaping approaches}

\subsubsection{$\tau$\/~leaping} \label{sec:tauLeaping}

As mentioned previously, the primary shortcoming of exact-stochastic simulation methods, whether applied to well-mixed systems or otherwise, is that every event firing is simulated explicitly.  This imposes a tremendous computational burden on the algorithm, particularly if one or more species have large populations.

To address this problem, Gillespie proposed the $\tau$\/-leaping approach, which proceeds by firing multiple reaction events at each simulation step \cite{Gillesp01}. In the well-mixed case, we first define the random variable $K_\mu(\tau)$ as the number of times reaction channel $R_\mu$\/ fires during the time interval $\left[t,t\!+\!\tau\right)$. The time evolution of the system can be formally written in terms of this variable as
\begin{equation}
    \mathbf{X}(t+\tau) = \mathbf{X}(t) + \sum_{\nu=1}^M \mathbf{z}_\nu K_\nu(\tau).
    \label{eq:X-t+tau}
\end{equation}
The idea then is to calculate some $\tau$\/ over which all reaction propensities remain ``essentially constant". In such a case, the reaction dynamics can be assumed to obey \emph{Poisson statistics\/} and
\begin{equation}
    K_\mu(\tau) \approx \mathcal{P}(a_\mu(t)\tau), \label{eq:K-Poiss}
\end{equation}
where $\mathcal{P}(a_\mu(t)\tau)$ is a Poisson random variable with mean and variance $a_\mu(t)\tau$.  Note that the dependence in Eq.~(\ref{eq:K-Poiss}) on the value of $a_\mu$\/ at the beginning of the step, i.e., at the initial time $t$\/, makes this an ``explicit" approach, analogous to explicit methods used in the numerical integration of ODEs \cite{Rath03}.

Equations~(\ref{eq:X-t+tau}) and (\ref{eq:K-Poiss}) constitute the essence of the (explicit) $\tau$\/-leaping method. At each step of a simulation, a time step $\tau$\/ is calculated (see Sec.~\ref{sec:tauSelec} below) and the system state updated by generating $M$\/ Poisson random deviates $\{k_\nu(\tau)\}$ in keeping with Eq.~(\ref{eq:K-Poiss}). Added to this is a \textit{proviso\/} that if the total number of expected firings, $a_0(t)\tau$, is ``small" ($\sim\!10$) then some variant of the \SSA{s} is used instead \cite{Gillesp01}.

Since its inception, modifications to the $\tau$\/-leaping approach have been proposed by various investigators  \cite{Gillesp03, Rath03, Cao05:negPop, Cao06:newStep, Cao07,Tian04, Chatt05, Auger06, Cai07, Petti07, Peng07, Rath07, Ander08, Xu08, Leier08, Harris06}. There are recent reviews by Gillespie~\cite{Gillesp07} and Pahle~\cite{Pahle09}. Though differing in various aspects, all of these methods are based on the same basic principles encapsulated in Eqs.~(\ref{eq:X-t+tau}) and (\ref{eq:K-Poiss}).

\subsubsection{Partitioned leaping} \label{sec:partLeaping}

In Refs.~\cite{Gillesp01} and \cite{Gillesp00}, Gillespie went beyond Eq.~(\ref{eq:K-Poiss}) and noted a well-known property of the Poisson distribution that it can be approximated by a \emph{normal\/}, or \emph{Gaussian\/}, distribution if the mean is ``large."  This allows us to write
\begin{align}
    K_\mu(\tau) & \approx \mathcal{P}(a_\mu(t)\tau) \approx \mathcal{N}(a_\mu(t)\tau,a_\mu(t)\tau) \nonumber \\     
    & = a_\mu(t)\tau + \sqrt{a_\mu(t)\tau} \times \mathcal{N}(0,1)
    \label{eq:K-Gauss}
\end{align}
where $\mathcal{N}(0,1)$ is a normal random variable with mean zero and unit variance \cite{Gillesp00, Gillesp01}.  Written this way, Eq.~(\ref{eq:K-Gauss}) is equivalent to the chemical Langevin equation \cite{Gillesp00}, a stochastic differential equation comprised of a ``deterministic" term and a fluctuating ``noise" term.  Gillespie then noted that as $a_\mu(t)\tau\!\rightarrow\!\infty$ the noise term becomes negligible relative to the deterministic term, giving
\begin{equation}
    K_{\mu}(\tau) \approx a_\mu(t)\tau, \label{eq:K-determ}
\end{equation}
which is equivalent to the forward-Euler method for solving deterministic ODEs \cite{Gillesp00, Gillesp01}.

In Ref.~\cite{Harris06}, we introduced the partitioned-leaping algorithm, a $\tau$\/-leaping variant that utilizes the entire theoretical framework encompassed by Eqs.~(\ref{eq:K-Poiss})--(\ref{eq:K-determ}).  The \PLA{l} considers reactions individually in a way reminiscent of the \NRM{s}. After calculating a time step $\tau$\/ (Sec.~\ref{sec:tauSelec}), each reaction is \emph{classified\/} into one of four categories: \ES{l}, Poisson, Langevin and deterministic. Reactions classified at the three coarsest levels (Poisson, Langevin, deterministic) utilize Eqs.~(\ref{eq:K-Poiss})--(\ref{eq:K-determ}), respectively. Reactions classified at the \ES{l} level are handled as in the \NRM{s} [Eqs.~(\ref{eq:FRM-tauMu}) and (\ref{eq:NRM-tauMu})].  Incorporating the \SSA{s} into the multiscale framework of the \PLA{l} is thus seamless and simple.  Details of the algorithm can be found in ref.~\cite{Harris06}, with a demonstration of its utility in ref.~\cite{Harris09}.%

\subsubsection{$\tau$\/~selection} \label{sec:tauSelec}

The central task in leaping algorithms is the manner in which the time step $\tau$\/ is determined.  Indeed, the entire method hinges on the validity of the Poisson approximation Eq.~(\ref{eq:K-Poiss}), which requires that the propensities of all reactions change negligibly during $\tau$\/.  To quantify this requirement, Gillespie defined the ``leap condition" \cite{Gillesp01, Gillesp00},
\begin{equation}
    \left|a_{\mu}(t+\tau) - a_{\mu}(t)\right| / \xi \leq \epsilon,
    \,\,\, (0 < \epsilon \ll 1)
    \label{eq:leapCondition}
\end{equation}
where $\xi$ is an appropriate scaling factor (see below).

Three main classes of $\tau$\/-selection procedure have been proposed: (i) a pre-leap \RB{l} (\RB{s}) approach that uses Eq.~(\ref{eq:leapCondition}) directly \cite{Gillesp01, Gillesp03, Cao06:newStep, Harris06}, (ii) a pre-leap \SB{l} (\SB{s}) approach where changes in the species populations are constrained such that Eq.~(\ref{eq:leapCondition}) is satisfied for all reactions \cite{Cao06:newStep, Harris06}, and (iii) a post-leap checking procedure that explicitly ensures that Eq.~(\ref{eq:leapCondition}) is satisfied at all simulation steps \cite{Ander08}.  Gillespie's initial $\tau$\/-selection strategy was an \RB{l} approach with $\xi\!\equiv\!a_0(t)$ \cite{Gillesp01, Gillesp03}, which we will refer to as RB-$a_0$.  More recently, Cao~et~al.~\cite{Cao06:newStep} proposed an improved \RB{l} approach with $\xi\!\equiv\!a_\mu(t)$, which we will refer to as RB-$a_\mu$\/, as well as a \SB{l} approach, which we will refer to as SB-$a_\mu$\/. The central task in this article involves modifying these formulas for use in spatial simulations (see Sec.~\ref{sec:spTauSelec}).

\subsubsection{Spatial $\tau$\/-leaping} \label{sec:spatialLeaping}

Spatial leaping approaches involve grouping events (reaction and diffusion) by subvolume, calculating a characteristic time interval $\tau_l^\mathrm{leap}$\/ for each subvolume and then choosing the global time step 
\begin{equation}
    \tau = \min_{l^\prime \in \{1{\ldots}L\}}\{\tau_{l^\prime}^\mathrm{leap}\}. \label{eq:spTauLeap} 
\end{equation}
Every reaction and diffusion event can then fire multiple times within $\tau$\/.

\MLB{full}~\cite{Lago07} attempted to generalize the \NSM{s} within the framework of such a leaping algorithm. The local time intervals $\tau_l^\mathrm{leap}$\/ are calculated using the RB-$a_0$ $\tau$\/-selection procedure of Gillespie and Petzold \cite{Gillesp03}, modified accordingly to apply to each subvolume. A binomial $\tau$\/-leaping variant \cite{Tian04} is used for calculating event firings and provisions are made to segue to the \NSM{s} when the species populations are small.

\RBK{full}~\cite{Rossi08} presented a similar implementation of spatial $\tau$\/-leaping with the primary difference being that they considered reaction and diffusion events independently of each other. Interestingly, they did not provide provisions to segue to a \SSA{s} method in the limit of small populations.

There are, however, some conceptual errors with both \MLB{fauthor}'s and \RBK{fauthor}'s spatial $\tau$\/-leaping methods. We aim address these concerns in Sec. \ref{sec:SPLA} in our development of the \SPLA{lcap} and outline the differences between the three spatial leaping algorithms in Sec. \ref{sec:variants}.

An important aspect of the spatial $\tau$\/-leaping algorithms is that, contrary to the exact-stochastic case (Sec.~\ref{sec:inhomogeneous}), grouping events by subvolume does not reduce the total number of calculations required in $\tau$\/~selection.  In the \NSM{s}, a characteristic time interval $\tau_l^\mathrm{exact}$ can be obtained for a given subvolume via a single evaluation of Eq.~(\ref{eq:FRM-tauMu}) with $a_\mu(t)$ replaced by $a_{l0}(t)$.  Thus, $L$\/ total calculations are required to determine $\tau$\/.  In the spatial $\tau$\/-leaping case, however, each $\tau_l^\mathrm{leap}$\/ requires performing $\tau$\/-selection calculations for each reaction (RB-$a_0$/RB-$a_\mu$\/) or species (SB-$a_\mu$\/) in $V_l$\/.  The total number of calculations required to determine $\tau$\/ in this context thus far exceeds $L$\/.  

This is a fundamental difference between the approaches that complicates the incorporation of spatial \SSA{s} methods like the \NSM{s} into a spatial leaping framework.  In Sec.~\ref{sec:spTauSelec}, we present optimized pre-leap $\tau$\/-selection formulas for subvolume-based spatial $\tau$\/-leaping methods that minimize computational effort by only considering those events that directly affect each reaction or species in $V_l$\/.  In Sec.~\ref{sec:discussion}, we speculate on alternative approaches that can fundamentally reduce the cost of $\tau$\/~selection by allowing a single calculation to be performed for a group of events, analogous to the procedure employed in the \NSM{s}.

\section{The spatial partitioned-leaping algorithm (SPLA)} \label{sec:SPLA}

\subsection{Motivation} \label{sec:motivation}

A major concern with the \MLB{full} method is the exclusion of incoming diffusion events in the $\tau$\/-selection process. In the \NSM{s}, incoming diffusion can be ignored when selecting values of $\tau$ because events outside of the subvolume have no bearing on when the next event within the subvolume will fire. In leaping methods, however, this is no longer the case: the relationships between events are of central importance in selecting values of $\tau$. Ignoring incoming diffusion in $\tau$\/~selection is thus an error that may impact the accuracy and/or efficiency of the method to an \emph{a priori} indeterminable extent. In Sec.~\ref{sec:examples}, we will show cases where this leads to inappropriately large values of $\tau$\/ and, hence, increased error, and cases where it results in unnecessarily small values of $\tau$\/ and decreased efficiency. Another concern in \MLB{fauthor}'s method is the use of the RB-$a_0$ $\tau$\/-selection procedure which is not as theoretically sound as (and has been shown to be less accurate than) the RB-$a_\mu$\/ and SB-$a_\mu$\/ procedures \cite{Cao06:newStep}. It appears that the RB-$a_0$ method was chosen to emulate the \NSM{s}.

In the case of \RBK{fauthor}\/'s method, the primary concern is the independent consideration of reactions and diffusion events during $\tau$\/~selection. In principle, this is inappropriate because the firings of reactions are intimately related to the rates at which entities diffuse into and out of subvolumes, and \textit{vice versa\/}. Ignoring this fact can introduce error and/or affect the efficiency of the method. Furthermore, the exclusion of a mechanism for transitioning to a \ES{l} method in the limit of small populations introduces additional error, as shown in Sec.~\ref{sec:examples}.

\subsection{Spatial $\tau$\/~selection} \label{sec:spTauSelec}

In \SPLA{s}, we address each of the above issues: (i) both incoming and outgoing diffusion are taken into account in the $\tau$\/-selection process, (ii) reactions and diffusion events are considered together when selecting time steps, (iii) appropriately modified formulations of the RB-$a_\mu$\/ and SB-$a_\mu$\/ $\tau$\/-selection procedures are used, and (iv) the method automatically segues to an \ES{l} method (NRM) at low populations.

In general, the \SPLA{s} can be seen as an accurate, straightforward implementation of spatial leaping against which future enhancements can be compared.  The method was not intended to be faster than other spatial $\tau$\/-leaping methods, though this is a worthy goal, and, as we shall see, it often is not faster. In such cases, the advantage of using SPLA should be measured in terms of accuracy. Sometimes SPLA is faster than other methods because it produces larger time steps.  This is particularly true for systems close to equilibrium where neglecting incoming diffusion can cause the algorithm to determine that the leap condition Eq.~(\ref{eq:leapCondition}) has been violated sooner than it actually has.

As in previous implementations of spatial $\tau$\/-leaping, we select time steps by calculating leap time intervals $\tau_l^\mathrm{leap}$ for each subvolume $V_l$\/ and then setting $\tau$\/ equal to the smallest of these [Eq.~(\ref{eq:spTauLeap})].  In Table~\ref{table:RBTauSelec}, we present a spatial version of the RB-$a_\mu$\/ $\tau$\/-selection procedure used in this article.  In Table~\ref{table:SBTauSelec}, we present the equations for the spatial SB-$a_\mu$\/ procedure. We pay special attention to the ranges over which  minimizations and summations are performed in these equations. In the RB-$a_\mu$\/ case, one value of $\tau_{l\mu}^\mathrm{leap}$ is calculated for each of the $M_l\!+\!N_l\Gamma_l$ reaction and \emph{outgoing\/} diffusion events in $V_l$\/. In the SB-$a_\mu$\/ procedure, one $T_{li}^\mathrm{leap}$ calculation is required for each of the $N_l$\/ species in $V_l$\/. In Eqs.~(\ref{eq:spRB-mLMu}), (\ref{eq:spRB-sigLMu}), (\ref{eq:spSB-mLI}) and (\ref{eq:spSB-sigLI}), summations are taken over \emph{all\/} $M_l\!+\!2N_l\Gamma_l$ events associated with $V_l$\/.  This is necessary to take into account the effect of \emph{incoming} diffusion and is critical for implementing an accurate spatial leaping algorithm.

\begin{table}[t]
\centering%
\caption{Spatial versions of the RB-$a_\mu$\/ $\tau$\/-selection formulas of Cao~et~al.~\cite{Cao06:newStep}, as modified in Harris and Clancy \cite{Harris06}.  One $\tau_{l\mu}^\mathrm{leap}$ calculation is required for each reaction and \emph{outgoing} diffusion event in $V_l$\/.   Note that in Eq.~(\ref{eq:spRBaMu-betaLMu}), $a_{l\mu}^\mathrm{min}$ is the smallest possible non-zero value of $a_{l\mu}$\/ ($a_{l\mu}^\mathrm{min}\!=\!c_{l\mu}$\/ for elementary reactions).}%
\begin{tabular}{c} \hline\hline
    Spatial RB-$a_\mu$ \\ \hline
    \begin{minipage}[t]{\columnwidth} \vspace{-12pt}
        \begin{gather}
            \tau_l^\mathrm{leap} = \min_{\nu \in \{1{\ldots}M_l+N_l\Gamma_l\}}\{\tau_{l\nu}^\mathrm{leap}\} \label{eq:spRB-tauL} \\[6pt]
            \tau_{l\mu}^\mathrm{leap} = \min\left\{\frac{\epsilon_{l\mu}(t)}{|m_{l\mu}(t)|} , \frac{\epsilon_{l\mu}^2(t)}{\sigma_{l\mu}^2(t)}\right\} 
                \label{eq:spRBaMu-tauLMu} \\[6pt]
            \epsilon_{l\mu}(t) \equiv \max\left\{\epsilon a_{l\mu}(t),\beta_{l\mu}(t)\right\} \label{eq:spRBaMu-epsLMu} \\[6pt]
            \beta_{l\mu}(t) =
                \begin{cases}
                    a_{l\mu}^{\min} \,\, \textit{if all\/} \,\, \left\{\frac{\partial a_{l\mu}(t)}{\partial X_{lj}} \right\} = 0 \\[6pt]
                    {\displaystyle \min_{\scriptscriptstyle j \in \{1{\ldots}N_l\}}}\!\left\{ \frac{\partial a_{l\mu}(t)}{\partial X_{lj}}\right\} 
                        \textit{\,\,otherwise} \\
                \end{cases} \label{eq:spRBaMu-betaLMu} \\[6pt]
            m_{l\mu}(t) \equiv \sum_{\nu=1}^{M_l+2N_l\Gamma_l} f_{l\mu\nu}(t) a_{l\nu}(t) \label{eq:spRB-mLMu} \\[6pt]
            \sigma_{l\mu}^2(t) \equiv \sum_{\nu=1}^{M_l+2N_l\Gamma_l} f_{l\mu\nu}^2(t) a_{l\nu}(t) \label{eq:spRB-sigLMu} \\[6pt]
            f_{l\mu\nu}(t) \equiv \sum_{j=1}^{N_l} z_{l\nu j} \frac{\partial a_{l\mu}(t)}{\partial X_{lj}} \label{eq:spRB-fLMuNu} 
        \end{gather}
    \end{minipage} \\[-6pt] 
    \\ \hline\hline
\end{tabular}
\label{table:RBTauSelec}
\end{table}

\begin{table}[t]
\centering%
\caption{Spatial versions of the SB-$a_\mu$\/ $\tau$\/-selection formulas of Cao~et~al.~\cite{Cao06:newStep}.  One $T_{li}^\mathrm{leap}$ calculation is required for each species in $V_l$\/.  Note that in Eq.~(\ref{eq:spSB-eLI}), the parameter $g_{li}$\/ depends on the types of events species $S_{li}$\/ participates in.  See \cite{Cao06:newStep} for formulas applicable to elementary event types, \cite{Harris06} for simplified versions of these, and \cite{Harris09} for extensions to select non-elementary events.}%
\begin{tabular}{c} \hline\hline
    Spatial SB-$a_\mu$ \\ \hline
    \begin{minipage}[t]{\columnwidth} \vspace{-12pt}
        \begin{gather}
            \tau_l^\mathrm{leap} = \min_{j \in \{1{\ldots}N_l\}}\{T_{lj}^\mathrm{leap}\} \label{eq:spSB-tauL} \\[6pt]
            T_{li}^\mathrm{leap} = \min\left\{\frac{e_{li}(t)}{|\widehat{m}_{li}(t)|} , \frac{e_{li}^2(t)}{\widehat{\sigma}_{li}^2(t)}\right\} 
                \label{eq:spSB-teeLI} \\[6pt]
            e_{li}(t) \equiv \max\left\{\epsilon X_{li}(t)/g_{li},1\right\} \label{eq:spSB-eLI} \\[-2pt]
            (0 < g_{li} < \infty) \nonumber \\[6pt]
            \widehat{m}_{li}(t) \equiv \sum_{\nu=1}^{M_l+2N_l\Gamma_l} z_{l\nu i} a_{l\nu}(t) \label{eq:spSB-mLI} \\[6pt]
            \widehat{\sigma}_{li}^2(t) \equiv \sum_{\nu=1}^{M_l+2N_l\Gamma_l} z_{l\nu i}^2 a_{l\nu}(t) \label{eq:spSB-sigLI}
        \end{gather}
    \end{minipage} \\[-6pt] 
    \\ \hline\hline
\end{tabular}%
\label{table:SBTauSelec}
\end{table}

\subsection{The algorithm} \label{sec:algo}

We define a domain of constant volume and divide it into $L$\/ (not necessarily equal-sized) subvolumes, each of volume $\omega_l$\/, using a finite difference type discretization. A connectivity matrix $\mbox{\boldmath$\mathsf{C}$}\!=\!\{\mathbf{C}_1,\ldots,\mathbf{C}_L\}$ is used to specify the neighboring subvolumes and the geometry of the domain. Boundary conditions are applied (e.g., periodic, reflecting) by appropriately defining {\boldmath$\mathsf{C}$}. The \SPLA{s} then proceeds as follows:

\begin{enumerate}

\item{\textit{Initialization\/}:
\begin{list}{\labelitemi}{\leftmargin=1.5em}
    \item[(i)]{For each subvolume, $V_l$\/: Set initial populations $\mathbf{X}_l(0)$ for all $N_l$\/ local species and define $M_l$\/ reactions in which these 
               species participate. Calculate initial values of the propensities $\{a_{l\nu}(0)\}$, $\nu\!\in\!\{1{\ldots}M_l\!+\!N_l\Gamma_l\}$ for all 
               reactions and \emph{outgoing\/} diffusion events. Set the time variable $t\!=\!t_\mathrm{init}$.}\\[-12pt]

    \item[(ii)]{Define global parameters $\epsilon$\/ (\mbox{$\ll\!1$}), \mbox{`$\approx\!1$'} and \mbox{`$\gg\!1$'} used in $\tau$\/~selection and event 
                classification (typical values are $0.01$--$0.05$, 3 and 100, respectively \cite{Harris06}).}
\end{list} \label{algo:initialize}}

\item{Calculate an initial (global) time step $\tau$\/ [Eq.~(\ref{eq:spTauLeap})] using either the RB-$a_\mu$\/ $\tau$\/-selection procedure of Table~\ref{table:RBTauSelec} or the SB-$a_\mu$\/ procedure of Table~\ref{table:SBTauSelec}. \label{algo:tauCalc}}

\item{Classify all $M_l\!+\!N_l\Gamma_l$ reaction and outgoing diffusion events within each $V_l$\/ based on the values of $a_{l\mu}(t)\tau$\/ (see Sec.~\ref{sec:partLeaping}). Prevent classification of diffusion events as \ES{l} if the population of the diffusing species $X_{li}(t)\!>\!100$ (see Sec.~\ref{sec:techIssues}). \label{algo:classify}}

\item{For all events (newly) classified as \ES{l}, generate values of $\tau_{l\mu}^\mathrm{exact}$ using Eqs.~(\ref{eq:FRM-tauMu}) and/or (\ref{eq:NRM-tauMu}).
      \label{algo:ESevents}}

\item{\begin{list}{\labelitemi}{\leftmargin=1.5em}
    \item[(i)]{If $\min\{\tau_{l^\prime\nu}^\mathrm{exact}\}\!<\!\tau$\/, $l^\prime\!\in\!\{1{\ldots}L\}$, $\nu\!\in\!\{$\textit{all \ES{l} events\/}$\}$, 
               set $\tau\!=\!\min\{\tau_{l^\prime\nu}^\mathrm{exact}\}$ and return to step~\ref{algo:classify} (this may require multiple iterations; 
               see \cite{Harris06}). \label{algo:reclassify}}\\[-12pt]

    \item[(ii)]{Else, if $\min\{\tau_{l^\prime\nu}^\mathrm{exact}\}\!>\!\tau$\/ \emph{and\/} all events are classified as \ES{l}, set 
                $\tau\!=\!\min\{\tau_{l^\prime\nu}^\mathrm{exact}\}$ (no iterations required).}\\[-12pt]

    \item[(iii)]{Else, retain $\tau$\/.}
\end{list} \label{algo:EStau}}

\item{Determine the numbers of event firings $\{k_{l^\prime\nu}(\tau)\}$, $l^\prime\in\{1{\ldots}L\}$, $\nu\in\{1{\ldots}M_l\!+\!N_l\Gamma_l\}$, based on the classifications. For the three coarsest descriptions, Eqs.~(\ref{eq:K-Poiss})--(\ref{eq:K-determ}) are used, respectively \cite{NumRec}. For \ES{l} events, if $\tau_{l\mu}^\mathrm{exact}\!=\!\tau$\/ then $k_{l\mu}(\tau)\!=\!1$, otherwise zero. \label{algo:firings}}

\item{Fire all events and update populations. \label{algo:update}} 

\item{If any $X_{li}(t\!+\!\tau)\!<\!0$, revert \emph{all\/} populations to their previous values, determine the numbers of event firings within the shorter
      time interval $[t\!+\!\tau/2)$ as $\{k_{l^\prime\nu}(\tau/2)\!=\!\mathcal{B}(k_{l^\prime\nu}(\tau),1/2)\}$, $l^\prime\in\{1{\ldots}L\}$, 
      $\nu\in\{1{\ldots}M_l\!+\!N_l\Gamma_l\}$, where $\mathcal{B}(n,p)$ is a binomial random deviate with $n$\/ attempts and a success probability of $p$\/ 
      (post-leap checking \cite{Ander08}; see Sec.~\ref{sec:techIssues}) and set $\tau\!=\!\tau/2$. Return to step~\ref{algo:update}. \label{algo:postLeap}}

\item{Advance the time to $t\!+\!\tau$ and return to step~\ref{algo:tauCalc} unless stopping criterion has been satisfied. \label{algo:lastStep}}

\end{enumerate}

\subsection{Technical issues} \label{sec:techIssues}

In step~\ref{algo:classify} of the \SPLA{s}, we include a provision that diffusion events should not be classified as \ES{l} if the populations of the diffusing species are greater than 100. This is a somewhat arbitrary restriction that deserves explanation.  In our initial trials, we often obtained time steps much smaller than expected, significantly diminishing the efficiency of the method, sometimes to a level close to that of the \NRM{s}.  We identified the source of this problem as diffusion events at the leading edge of diffusing fronts.  In these regions, the numbers of diffusing molecules are small and, as such, diffusion events obtain \ES{l} classifications.  In many instances, the values of $\tau_{l\mu}^\mathrm{exact}$ generated for these events were smaller than $\tau$\/, requiring a reduction in the time step and a reclassification of all events [step~\ref{algo:reclassify}(i) above].  This often led to events in subvolumes away from the leading edge being classified as \ES{l} that previously were not, which would then produce an even smaller time step, and so.  This ``classification cascade" ultimately resulted in values of $\tau$\/ much smaller than necessary.  The same behavior was observed in a previous application of the \PLA{s} to a model biological system \cite[note~80]{Harris09}.

The provision in step~\ref{algo:classify} of the \SPLA{s} was included in order to overcome this problem.  It prevents the cascade from penetrating too deep into the interior of the domain and significantly speeds the simulations with negligible loss in accuracy.  Our choice of 100 as the threshold is based on the fact that diffusion is usually modeled as a first-order process and, hence, if the population is 100 then one firing will result in a 1\% change in the propensity.  1\% is a reasonable value for $\epsilon$\/ and is at the lower end of the typical values that we use.  Nevertheless, this approach is clearly \textit{ad~hoc\/} and it would be preferable to have a more general strategy that applies globally to all event types, not just diffusion events. In the future, we hope to develop such an approach. For the sake of demonstration, however, we believe that this simple strategy suffices.

In step~\ref{algo:postLeap} of the \SPLA{s}, we employ the post-leap checking procedure of Anderson \cite{Ander08}, which is theoretically stronger than the ``try again" approach employed in (step~8 of) the original \PLA{s}~\cite{Harris06} and in $\tau$\/-leaping \cite{Cao05:negPop, Cao06:newStep}. However, would like to emphasize that in the \SPLA{s}, we make minimal use post-leap checking and only to handle those \emph{rare} occasions in which negative populations arise. Post-leap checking has much broader potential as an alternative $\tau$\/-selection approach that can improve the efficiency of the \SPLA{s}, either on its own or coupled with the \RB{l} or \SB{l} procedures of Tables~\ref{table:RBTauSelec} and \ref{table:SBTauSelec}.

\subsection{Marquez-Lago, Rossinelli and some SPLA variants} \label{sec:variants}

In order to assess the performance of the \SPLA{s}, we implemented \MLB{fauthor}'s and \RBK{fauthor}'s spatial $\tau$\/-leaping methods for comparison, as well as variants of the \SPLA{s} that incorporate select features of those methods for diagnostic purposes.  \MLB{fauthor}'s method differs from the \SPLA{s} in two important ways: (i) it calculates values of $\tau_{l\mu}^\mathrm{leap}$ using the RB-$a_0$ $\tau$\/-selection procedure 
\begin{equation}
    \tau_{l\mu}^\mathrm{leap} = \min\left\{\frac{\epsilon a_{l0}(t)}{|m_{l\mu}(t)|} , \frac{\epsilon^2 a_{l0}^2(t)}{\sigma_{l\mu}^2(t)}\right\},
    \label{eq:spRBa0-tauMu}
\end{equation}
where $a_{l0}(t)\!\equiv\!\sum_{\nu=1}^{M_l+N_l\Gamma_l} a_{l\nu}(t)$, and (ii) incoming diffusion is ignored in these calculations.  The latter means that $m_{l\mu}(t)$ and $\sigma_{l\mu}^2(t)$ are calculated as in Eqs.~(\ref{eq:spRB-mLMu}) and (\ref{eq:spRB-sigLMu}) of Table~\ref{table:RBTauSelec} but with the summations running over $\nu\!\in\!\{1{\ldots}M_l\!+\!N_l\Gamma_l\}$ only.  Values of $\tau_l^\mathrm{leap}$ are calculated using Eq.~(\ref{eq:spRB-tauL}) of Table~\ref{table:RBTauSelec} and $\tau$\/ is selected as in Eq.~(\ref{eq:spTauLeap}). 

\MLB{fauthor}'s method also transitions to using an \ES{l} method in $V_l$\/ if $a_{l0}(t)\tau\!\lesssim\!10$.  This amounts to classifying the \emph{subvolume\/} as \ES{l} which, in turn, experiences either one event firing within $\tau$\/ or none at all.  If one event fires, then event selection is performed as in the \DM{l}. Consequently, if all subvolumes are classified as \ES{l}, then the algorithm \emph{becomes\/} the \NSM{s}~\cite{Elf04, Hattne05}. The numbers of firings within non-\ES{l} subvolumes are determined using a binomial $\tau$\/-leaping variant \cite{Tian04}. Importantly, the method does not use the continuum descriptions Eqs.~(\ref{eq:K-Gauss}) and (\ref{eq:K-determ}) that are used in the \SPLA{s}.  Note that, instead of binomial $\tau$\/-leaping, we use standard Poisson $\tau$\/-leaping coupled with the negative population check of step~\ref{algo:postLeap} of the \SPLA{s}. We consider this difference to be inconsequential in comparing the methods.

The primary differences between the \SPLA{s} and the \RBK{fauthor}'s spatial $\tau$\/-leaping method are: (i) they apply the SB-$a_\mu$\/ $\tau$\/-selection procedure of Table~\ref{table:SBTauSelec} separately to reaction and diffusion events and, (ii) they do not provide a mechanism for segueing to a \ES{l} method in the limit of small populations.  For each subvolume $V_l$\/, $\tau_l^\mathrm{leap}$ values are calculated by
\begin{equation}
    \tau_l^\mathrm{leap} = \min\{\tau_l^\mathrm{rxn}, \tau_l^\mathrm{diff}\}. \label{eq:RBK-tauL} 
\end{equation}
with $\tau_l^\mathrm{rxn}$ and $\tau_l^\mathrm{diff}$ being time steps for reactions and diffusion events respectively. They are obtained using modified forms of Eq.~(\ref{eq:spSB-tauL}) in Table~\ref{table:SBTauSelec}.  Basically, for each $S_{li}$\/, \emph{two\/} values of $T_{li}^\mathrm{leap}$ are calculated, one considering only reactions and the other only diffusion events (outgoing and incoming).  These are obtained via Eq.~(\ref{eq:spSB-teeLI}) of Table~\ref{table:SBTauSelec} with $\widehat{m}_{li}(t)$ and $\widehat{\sigma}_{li}^2(t)$ calculated using Eqs.~(\ref{eq:spSB-mLI}) and (\ref{eq:spSB-sigLI}), respectively, but with the summations running only over $\nu\!\in\!\{1{\ldots}M_l\}$ for $\tau_l^\mathrm{rxn}$ and $\nu\!\in\!\{M_l\!+\!1{\ldots}M_l\!+\!2N_l\Gamma_l\}$ for $\tau_l^\mathrm{diff}$.  Thus, in our implementation of \RBK{fauthor}'s method, we replace step~\ref{algo:tauCalc} of the \SPLA{s} with this $\tau$\/-selection procedure.  We also eliminate steps~\ref{algo:classify}--\ref{algo:EStau} of the \SPLA{s} and use only Eq.~(\ref{eq:K-Poiss}) in step~\ref{algo:firings} (i.e., no \ES{l}, Langevin or deterministic classifications).

Finally, we also implement three variants of the \SPLA{s}: (i) a ``one-way diffusion" variant that sums only over $\nu\!\in\!\{1{\ldots}M_l\!+\!N_l\Gamma_l\}$ in Eqs.~(\ref{eq:spRB-mLMu}) and (\ref{eq:spRB-sigLMu}) of Table~\ref{table:RBTauSelec} and Eqs.~(\ref{eq:spSB-mLI}) and (\ref{eq:spSB-sigLI}) of Table~\ref{table:SBTauSelec} during $\tau$\/~selection [step~\ref{algo:tauCalc} of the SPLA], (ii) a ``no ES reactions" variant that prevents reaction events from being classified as \ES{l} in step~\ref{algo:classify} of the SPLA, and (iii) a ``no ES events" variant that prevents all events (reaction and diffusion) from being classified as \ES{l}.  The first variant allows us to quantify the effects of ignoring incoming diffusion in $\tau$\/~selection. The last variant gives us insight into the importance or tradeoff of transitioning to a \ES{l} method in the limit of small populations. The second variant is used to exemplify the need for a more general strategy to address the classification cascade problem discussed in Sec.~\ref{sec:techIssues}. These variants provided us with insight into the operation of the \SPLA{s} and allowed us to make connections to  \MLB{fauthor}'s and \RBK{fauthor}'s methods.

\section{Numerical Examples} \label{sec:examples}

In order to demonstrate the utility of the \SPLA{s}, we apply the method to three classical spatial systems: pure diffusion in one dimension (Sec.~\ref{sec:pureDiffusion}), the one-component reaction-diffusion system described by Fisher's equation \cite{Fisher:1937, Kolmogorov:1937} in one dimension (Sec.~\ref{sec:FisherEq}), and the two-component reaction-diffusion system described by the Gray-Scott equations \cite{Pearson:1993} in two dimensions (Sec.~\ref{sec:GrayScott}).  In all cases, we consider the domain partitioned into $L$\/ equally-sized subvolumes.  Diffusion is modeled as a first-order elementary process
\begin{equation}
    S_{li} \xrightarrow{d_i} S_{l^\prime i}, \label{rxn:1stOrderDiff}
\end{equation}
where $V_{l^\prime}$\/ is an adjacent subvolume (i.e., $V_{l^\prime}\!\in\!\mathbf{C}_l$\/) and the microscopic diffusivity $d_i$\/ is constant throughout the domain. Propensities for diffusion events are thus of the form
\begin{equation}
    a_{l\mu}(t) = d_i X_{li}(t), \,\,\, \mu\in\{M_l+1{\ldots}M_l+N_l\Gamma_l\}. \label{eq:aMu-diff}
\end{equation}
Microscopic diffusivities are obtained from macroscopic diffusion coefficients $D_i$\/ via the relation \cite{Bernst05}
\begin{equation}
    d_i = D_i/h^2, \label{eq:microD}
\end{equation}
where $h$\/ is the side length of the regular subvolumes. We choose $h$ such that the size of the subvolume is less than the diffusion length of the system, given by $\sqrt{4D\tau}$ (where $D$ is diffusivity and $\tau$ is the time step). However, the time step $\tau$ can vary significantly during the course of the simulation. It is affected by the rate of diffusion, which in turn is affected by the subvolume size (Ref. Eq. (\ref{eq:microD})). Hence this formula can only be used approximately. This circular dependency can be partially addressed by running a sample simulation, taking the most-frequent time step and then using that to calculate the subvolume size such that the well-mixed assumption is maintained. All \SPLA{s} simulations are performed with $\epsilon\!=\!0.01$, `$\approx\!1$'$=\!3$ and `$\gg\!1$'$=\!100$.

\subsection{Pure diffusion} \label{sec:pureDiffusion}

The first system we considered was pure diffusion of a $\delta$ function in one dimension.  Apart from being the simplest example of a diffusing front, this system is ideal for study because analytical solutions are well known and the stochastic mean corresponds to the deterministic solution.

\begin{figure}
\centering%
\includegraphics{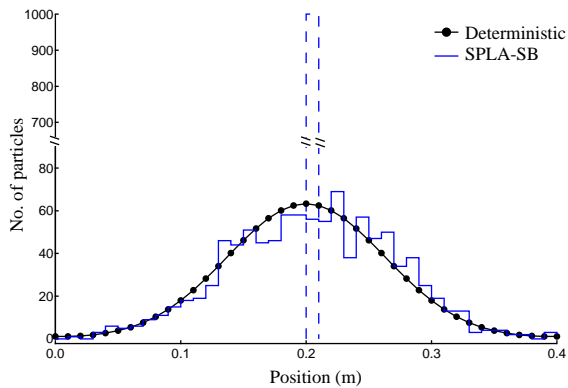} 
\caption{Particle distributions at $t\!=\!2$~s for pure diffusion of a 1000 particle $\delta$ spike obtained using the SPLA and Eq.~(\ref{eq:pureDiffPDE}). The initial delta function is shown as a dashed line. The cross-sectional area $A$\/ is set such that the total concentration over the domain is $0.04~M$\/. The SPLA simulation was performed using the SB-$a_\mu$\/ $\tau$\/-selection procedure of Table~\ref{table:SBTauSelec} (SPLA-SB).}
\label{fig:diffusion_schematic}
\end{figure}

We define a one-dimensional domain of width 0.4~m (in say, the \textit{y}-direction) and cross-sectional area $A$\/ and divide it into $L\!=\!40$ equally-sized subvolumes, each of width 0.01~m ($\omega_l\!=\!0.01A$~m$^3$). We populate one subvolume at the center of the domain [see Fig.~\ref{fig:diffusion_schematic}] with between $X(0)\!=\!1$ and $5\!\times\!10^7$ particles of species $S$\/ and then vary $A$\/ in order to maintain a constant concentration of $0.04~M$\/ over the whole domain. We apply Neumann (no flux) boundary conditions at each end of the domain, and define a constant (\textit{y}-directional) diffusion coefficient $D\!=\!10^{-3}$~cm$^2$/s. The system can then be represented by the set of transformations
\begin{gather}
    S_l \xrightleftharpoons[d]{d} S_{l+1}, \,\,\,\,\, l\in\{1{\ldots}L-1\}, \label{diff:pureDiff}
\end{gather}
where $d$\/ is obtained from Eq.~(\ref{eq:microD}).  The partial differential equation that describes this system in the deterministic limit is
\begin{equation}
    \frac{\partial X(y,t)}{\partial t} = D \frac{\partial^2 X(y,t)}{\partial y^2}. \label{eq:pureDiffPDE}
\end{equation}
In Fig.~\ref{fig:diffusion_schematic}, we compare particle distributions at $t\!=\!2$~s for an initial $\delta$ spike of 1000 particles obtained from a representative \SPLA{s} simulation of (\ref{diff:pureDiff}) and from Eq.~(\ref{eq:pureDiffPDE}).  The results coincide well, although the effects of stochasticity are clearly visible. 

\begin{figure}
\centering %
\includegraphics{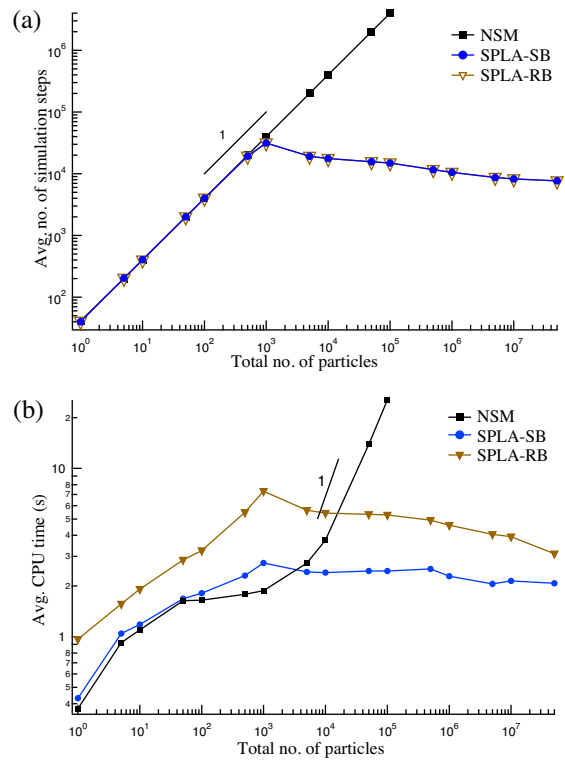}
\caption{(a) Average numbers of simulation steps and (b) average CPU times vs.\ total particle number for pure diffusion of a $\delta$ function till $t\!=\!2$~s using the SPLA-RB, SPLA-SB and NSM.  In each case, the particle number was changed by varying the cross-sectional area $A$, while maintaining a constant concentration of $0.04~M$\/ over the domain.  All results are averaged over 500 simulation runs performed on an Intel Core~2 Duo, 2.13~GHz machine with 2~GB of RAM.}
\label{fig:diffusion_time}
\end{figure}

In Fig.~\ref{fig:diffusion_time}, we present a computational cost analysis comparing the \SPLA{s} to the \NSM{s}.  In Fig.~\ref{fig:diffusion_time}(a), we see that the \SPLA{s}, using both the \RB{l} (\SPLA{s}-\RB{s}) and \SB{l} (\SPLA{s}-\SB{s}) $\tau$\/-selection procedures of Tables~\ref{table:RBTauSelec} and \ref{table:SBTauSelec}, requires almost exactly the same number of simulation steps as the \NSM{s} up to about 1000 total particles.  Beyond that, we see a significant difference between the methods, with the cost of the \SPLA{s} \emph{decreasing\/} with increasing number of particles and that of the \NSM{s} continuing to increase linearly.  The reason why the two \SPLA{s} methods coincide \emph{exactly\/} is because we model diffusion as a first-order elementary process [Eq.~(\ref{rxn:1stOrderDiff})].  Thus, the constraint on $|\Delta a_{l\mu}(t)|$ used in \RB{l} $\tau$\/~selection is identical to that on $|\Delta X_{li}(t)|$ used in \SB{l} $\tau$\/~selection.

In Fig.~\ref{fig:diffusion_time}(b), we compare the CPU times for each of the three methods. Here, we see that up to about 1000 total particles the \NSM{s} is actually the least expensive of the methods.  The \SPLA{s}-\SB{s} is close behind, however, being slightly less efficient because of the computational overhead associated with $\tau$\/~selection. Beyond 1000 total particles, we see that the \SPLA{s} decreases in computational cost while the cost of the \NSM{s} continues to increase linearly. Interestingly, \SPLA{s}-\RB{s} is significantly less efficient than the \SPLA{s}-\SB{s}, despite the fact that both methods take the exact same number of steps on average.  This is due to two factors: (i) the total number of $\tau^\mathrm{leap}$ calculations required in \RB{l} $\tau$\/-selection ($M_l\!+\!N_l\Gamma_l\!=\!78$) as compared to \SB{l} ($N_l\!=40$), and (ii) the extra expense associated with calculating rate derivatives in \RB{s} $\tau$\/-selection [Eqn. \ref{eq:spRB-fLMuNu}].  Since $N_l$\/ will often be much less than $M_l\!+\!N_l\Gamma_l$, we see that there is a distinct advantage to using \SB{s} $\tau$\/-selection in spatial leaping simulations.

\begin{figure}
\centering
\includegraphics{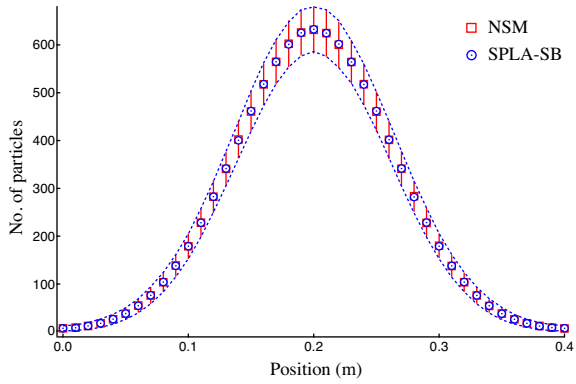} 
\caption{Means and standard deviations of the particle number over the entire domain at $t\!=\!2$~s for pure diffusion of a $10^4$ particle $\delta$ function using the SPLA-SB and the NSM.  In both cases, results are from 500 simulation runs.  The dotted lines constitute an envelope of twice the standard deviation about the SPLA-SB mean.}
\label{fig:diffusion_accuracy}
\end{figure}

In Fig.~\ref{fig:diffusion_accuracy}, we compare the accuracy of the \SPLA{s}-\SB{s} to the \NSM{s} for an initial $\delta$ spike of $10^4$ particles. We omit the \SPLA{s}-\RB{s} since the results are identical to the \SPLA{s}-\SB{s}. We see that, although the \SPLA{s} requires about an order of magnitude fewer steps [Fig.~\ref{fig:diffusion_time}(a)], there is essentially no difference between the means and standard deviations obtained from both methods over the entirety of the domain. We make sense of this by referring to the works of Cao et al. \cite{Cao04} and Rathinam et al. \cite{Rath05}, both of which show that in explicit $\tau$\/-leaping methods (like \SPLA{s}), for sufficiently small $\tau$, the histograms generated using a $\tau$-leaping method should be virtually indistinguishable from those obtained using an \ES{l} method. Our results in Fig.~\ref{fig:diffusion_accuracy} thus simply indicate that we are using a small enough error control parameter ($\epsilon\!=\!0.01$) in $\tau$\/~selection and thus avoiding any noticeable errors.

\subsection{Fisher's equation} \label{sec:FisherEq}

Fisher's equation (also known as the Fisher-Kolmogorov-Petrovskii-Piscounov equation) \cite{Fisher:1937, Kolmogorov:1937} is a deterministic partial differential equation that has been used to describe the propagation of an advantageous gene in a population \cite{Fisher:1937} and the spatio-temporal evolution of a species under the combined effects of diffusion and logistical growth \cite{Kolmogorov:1937}.  In one dimension, the equation is of the form 
\begin{equation}
    \frac{\partial u}{\partial t} = Ku(\bar{c}-u) + D\frac{\partial^2 u}{\partial y^2}, \label{eq:Fisher}
\end{equation}
where $u$\/ is a species concentration, $K$\/ is a second-order reaction rate constant, $\bar{c}$\/ is the ``carrying capacity" or ``saturation value" of the system and $D$\/ is a diffusion coefficient.

We again consider a one-dimensional domain of width 0.4~m and cross-sectional area $A$ and divide it into $L\!=\!40$ equally-sized subvolumes ($\omega_l\!=\!0.01A$~m$^3$) with Neumann (no flux) boundary conditions applied at each end.  On this domain, we consider the reaction-diffusion system
\begin{alignat}{2}
    S_{l1}& + S_{l2} \xrightarrow{k} 2S_{l1}, &\quad& l\in\{1{\ldots}L\}, \notag \\
    S_{l1}& \xrightleftharpoons[d]{d} S_{(l+1)1}, && l\in\{1{\ldots}L-1\}, \label{rxndiff:Fisher} \\
    S_{l2}& \xrightleftharpoons[d]{d} S_{(l+1)2}, && l\in\{1{\ldots}L-1\}. \notag
\end{alignat}
Because $S_1$\/ and $S_2$\/ have equal diffusivities throughout the domain, in the deterministic limit the total population within each subvolume $X_{lT}\!=\!X_{l1}(t)\!+\!X_{l2}(t)$\/ is constant.  The spatio-temporal evolution of $S_1$\/ can thus be described by Fisher's equation (\ref{eq:Fisher}) with $d\!=\!D/h^2$, $u\!=\!X_1(y,t)/\Omega$, $K\!=\!k\Omega$\/, $\bar{c}\!=\!X_{lT}/\Omega$, where $\Omega\!=\!N_A\omega_l$ and $N_A$\/ is Avogadro's number.  

\begin{figure}
\centering
\includegraphics{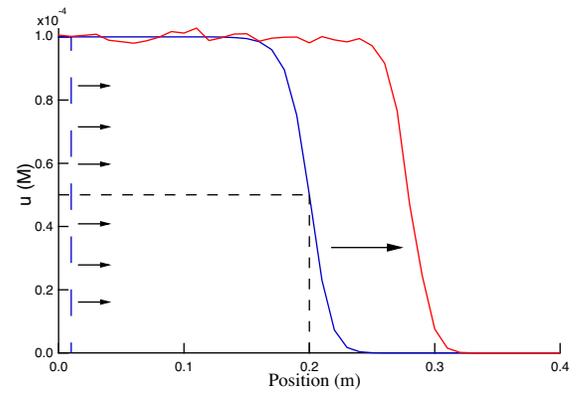}
\caption{Solution of the one-dimensional Fisher's equation (\ref{eq:Fisher}).  The initial condition is shown by the dashed lines. The deterministic  trajectory (blue) is shown at $t\!=\!3.73$~s, the time at which the solution reaches half its saturation value at $y\!=\!0.2$~m. A stochastic trajectory (red) is shown at $t\!=\!5.0$~s}
\label{fig:fisher_schematic}
\end{figure}

Initially, we take the first compartment to be saturated with $S_1$\/ [i.e, $X_{11}(0)\!=\!\bar{c}\Omega$\/; see Fig.~\ref{fig:fisher_schematic}] and all other compartments to be saturated with $S_2$\/. The saturation value $\bar{c}$\/ is taken be $10^{-4}~M$\/ and we choose $D\!=\!10^{-4}$~m$^2$/s and $K\!=\!7\!\times\!10^{4}~M^{-1}$~s$^{-1}$.  To investigate the effects of stochasticity, we hold $\bar{c}$\/ constant and vary the particle number $X_{lT}$ by varying the cross-sectional area $A$\/.  In Fig.~\ref{fig:fisher_schematic}, we show a snapshot of the traveling wave of $S_1$ obtained by solving Fisher's equation (\ref{eq:Fisher}) and the corresponding reaction-diffusion system (\ref{rxndiff:Fisher}).

\begin{figure}
\centering %
\includegraphics{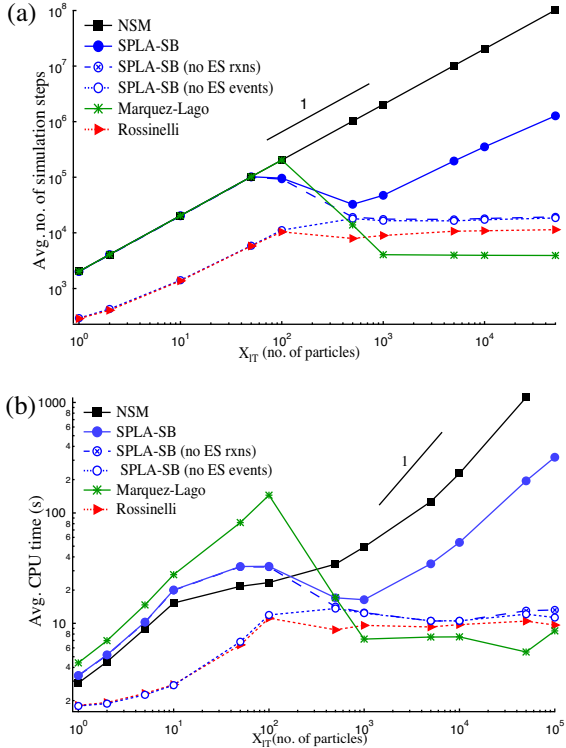}
\caption{(a) Average numbers of simulation steps and (b) average CPU times vs.\ $X_{lT}$\/ for simulations till $t\!=\!25$~s of the reaction-diffusion system (\ref{rxndiff:Fisher}) using various methods.  In each case, the particle number is changed by varying the cross-sectional area $A$ while maintaining a constant concentration of $\bar{c}$\/ within each $V_l$\/.  All results are averaged over 500 simulation runs performed on an Intel Core~2 Duo, 2.13~GHz machine with 2~GB of RAM.}
\label{fig:fisher_time}
\end{figure}

In Fig.~\ref{fig:fisher_time}, we show a computational cost analysis comparing different variants of the \SPLA{s}-\SB{s} to the \NSM{s} and \MLB{fauthor}'s and \RBK{fauthor}'s spatial $\tau$\/-leaping methods. [We do not consider the \SPLA{s}-\RB{s} since it is significantly less efficient than the \SPLA{s}-\SB{s}]. Simulations are run until $t\!=\!25$~s, and the results are averaged over 500 runs. In Fig.~\ref{fig:fisher_time}(a), we see that, at low populations, the numbers of simulation steps for all methods scale linearly with the number of particles, although the \RBK{fauthor} method and SPLA-(no \ES{s} events) take an order of magnitude fewer steps. This indicates that these methods are firing multiple events even when the populations are small.  Above about $X_{lT}\!=\!100$, however, we see a divergence from the linear trend for all of the leaping methods.  The cost of the \MLB{fauthor} and \RBK{fauthor} methods are independent of system size above this point, while that for the full \SPLA{s} drops initially, but then continues to increase linearly beyond about $X_{lT}\!=\!500$. However, when we selectively disable the \ES{l} classification for \emph{reactions only\/} in the \SPLA{s} we see that the cost decreases significantly, approaching those of \MLB{fauthor} and \RBK{fauthor}.  This indicates that, for this system, reaction events are causing a classification cascade at large populations just as diffusion events did in our initial studies.  This exemplifies the need to develop a more generalized approach for handling the classification cascade problem (see Sec.~\ref{sec:techIssues}). In Fig.~\ref{fig:fisher_time}(b), we see similar trends for the CPU times, although \MLB{fauthor}'s method and the \SPLA{s} are somewhat more costly than the \NSM{s} at small populations because of the added overhead associated with $\tau$\/~selection.

\begin{figure}
\centering
\includegraphics{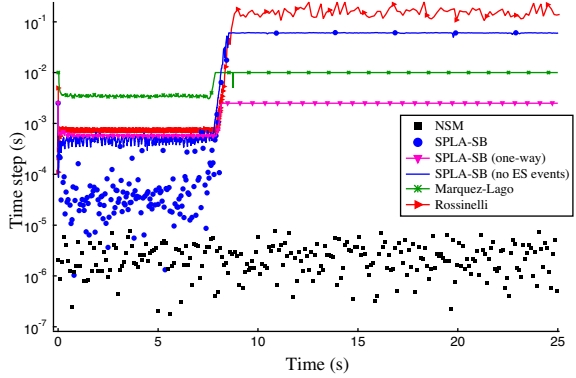}
\caption{Time steps calculated during individual simulation runs of the reaction-diffusion system (\ref{rxndiff:Fisher}) with $X_{lT}\!=\!10^4$ using various methods.}
\label{fig:fisher_tau}
\end{figure}

The results in Fig.~\ref{fig:fisher_time} would seem to indicate that the \SPLA{s} is always slower than both the \MLB{fauthor} and \RBK{fauthor} methods.  However, this is not entirely true. In Fig.~\ref{fig:fisher_tau}, we show the time steps taken during representative simulation runs with $X_{lT}\!=\!10^4$ for the various methods. We see that during the first $\sim\!7$~s, when the wave is propagating across the domain, the time steps for the \SPLA{s} are small.  \RBK{fauthor}'s method takes slightly larger time steps during this period while \MLB{fauthor}'s time steps are significantly larger.  The scatter of particularly small time steps for the full \SPLA{s} exemplifies the classification cascade effect evident in Fig.~\ref{fig:fisher_time}.  We also see how forbidding the \ES{l} classification for reactions prevents this from occurring. At $\sim\!7$~s, however, the situation changes dramatically.  The system approaches equilibrium and the time steps for all leaping methods increase significantly, with \RBK{fauthor}'s method experiencing the largest jump, followed by the \SPLA{s} and then \MLB{fauthor}.  Also note how the classification cascade problem ceases in the \SPLA{s}.

We make sense of these results by considering the one-way diffusion variant of the \SPLA{s} (Sec.~\ref{sec:variants}), where incoming diffusion is ignored in $\tau$\/~selection. We see in Fig.~\ref{fig:fisher_tau} that this results in significantly smaller time steps for $t\!\gtrsim\!7$~s. The time period after $\sim\!7$~s corresponds to the equilibrium state of the system, i.e., it takes $\sim\!7$~s for the wave to travel across the domain. At equilibrium, incoming diffusion replenishes the numbers of particles in subvolumes. Ignoring this causes the algorithm to underestimate the time at which the leap condition Eq.~(\ref{eq:leapCondition}) will be violated. This explains why the time steps for \MLB{fauthor}'s method are smaller during this phase than other leaping methods and, if we were to run the simulations longer than $25$~s, \SPLA{s} would become more efficient. The remaining disparity between \MLB{fauthor} and the one-way \SPLA{s} is due to differences in $\tau$\/-selection procedure. The larger time steps for \RBK{fauthor} during this phase are due to their separate consideration of reaction and diffusion events.

\begin{figure}
\centering
\includegraphics{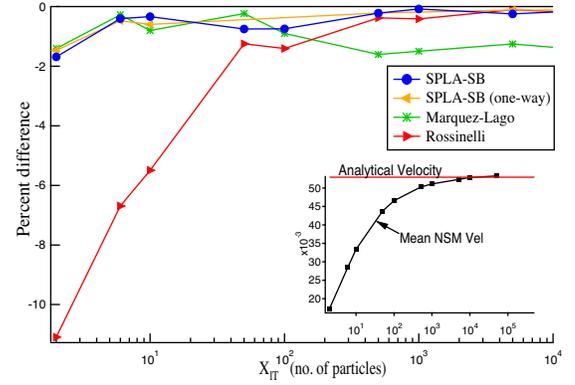}
\caption{Percent deviations between mean wave velocities obtained from various leaping methods and the NSM for varying system sizes ($\left\langle\mathcal{V}\right\rangle_\mathrm{leap}\!-\!\left\langle\mathcal{V}\right\rangle_\mathrm{NSM}/\left\langle\mathcal{V}\right\rangle_\mathrm{NSM}\!\times\!100\%$). (Inset) Convergence of $\left\langle\mathcal{V}\right\rangle_\mathrm{NSM}$ to the analytical solution $\mathcal{V}\!=\!2\sqrt{DK\bar{c}}$ \cite{Mai:2000} with increasing system size.  All results are based on 500 leaping and NSM simulation runs.  Note that the apparent discrepancies between the NSM and the SPLA and Marquez-Lago methods at small particle numbers are simply due to random sampling error [also see Fig.~\ref{fig:fisher_kolmogorov}(c)].}
\label{fig:fisher_velocity}
\end{figure}

We find that the different time steps obtained by various methods give rise to different traveling wave velocities $\mathcal{V}$\/, which we can use to compare the accuracies of the various methods. We measure the velocity as the time taken for $S_1$\/ to reach half its saturation value at $y\!=\!0.2$~m.  For the Heaviside initial condition, the analytical expression for the wave velocity is $\mathcal{V}\!=\!2\sqrt{DK\bar{c}}$ \cite{Mai:2000}.  However, stochastic effects give rise to a distribution of wave velocities for the same initial condition. Recent authors have shown that, depending on the values of $\bar{c}$ and $K$\/, the mean of the velocity distribution can differ from the analytical velocity \cite{Lemarchand:1995, Mai:1998, Panja:2004}, particularly at low populations. Thus, instead of the analytical solution, we use the mean velocity $\left\langle\mathcal{V}\right\rangle_\mathrm{\NSM{s}}$ obtained from $500$ \NSM{s} simulations as the standard for comparison. In Fig.~\ref{fig:fisher_velocity}, we show percent deviations between the mean wave velocities obtained from the various leaping methods and $\left\langle\mathcal{V}\right\rangle_\mathrm{\NSM{s}}$ as a function of $X_{lT}$.  In the inset, we show the convergence of $\left\langle\mathcal{V}\right\rangle_\mathrm{\NSM{s}}$ to the analytical solution with increasing number of particles.

Fig.~\ref{fig:fisher_velocity} shows that, for small populations, \RBK{fauthor}'s method has large errors in its wave velocity. The error decreases with increasing population and becomes negligible at the largest system sizes considered.  \MLB{fauthor}'s method, on the other hand, shows the opposite trend: the error is negligible at small populations and increases with increasing population.  We can explain these observations by referring back to Figs.~\ref{fig:fisher_time} and \ref{fig:fisher_tau}.  The error in \RBK{fauthor} at small populations is mainly due to the fact that the method lacks a mechanism for transitioning to a SSA method. Thus, the increase in efficiency seen in Fig.~\ref{fig:fisher_time} comes at the cost of accuracy. The error in \MLB{fauthor}'s method is due to the combined effect of the $\tau$-selection procedure and neglecting incoming diffusion. At small populations, the method transitions to \NSM{s}, thus reducing error.  At large populations, however, the method takes larger time steps than the other leaping methods during the wave-propagation phase (Fig.~\ref{fig:fisher_tau}), resulting in the increased error seen in Fig.~\ref{fig:fisher_velocity}. Furthermore, in the equilibrium phase (after $\sim\!7$~s), the method takes smaller steps than \SPLA{s} and the error then changes from one of accuracy to one of efficiency. \SPLA{s} addresses each of these issues and shows negligible error over the entire population range. SPLA-(one way) tries to capture just the effect of neglecting incoming diffusion. However, we do not observe any significant error because the method transitions to an \ES{l} method at small populations and, calculates leap time steps that are fairly near the accurate \SPLA{s} time steps at higher populations. 

\begin{figure*}
\centering
\includegraphics[width=500pt]{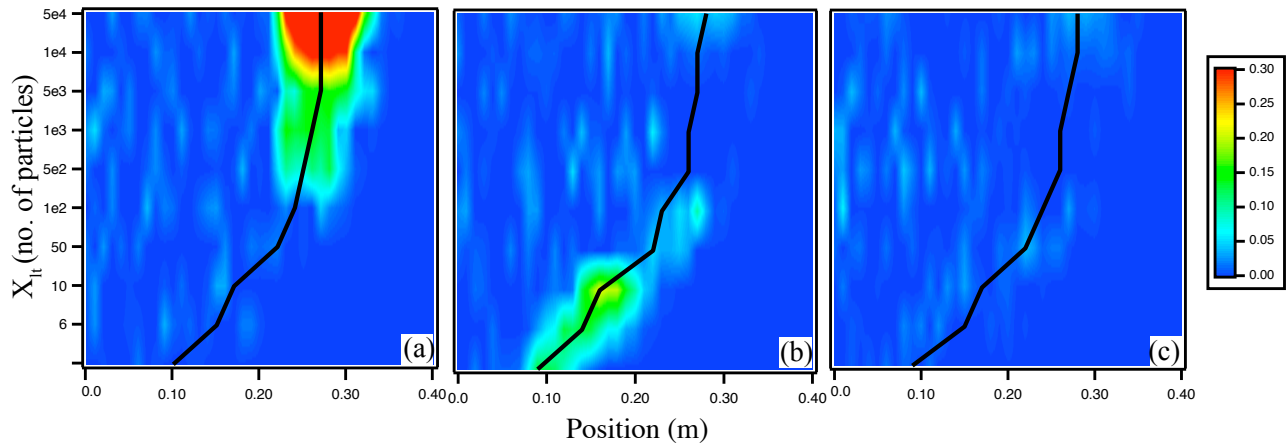}
\caption{Kolmogorov distances at $t\!=\!5$~s obtained using: (a) Marquez-Lago, (b) Rossinelli, and (c) SPLA. The x-axis corresponds to position within the domain and the y-axis to system size (i.e., $X_{lT}$\/).  All results are based on 500 leaping and NSM simulation runs.  The NSM self distance is subtracted from the Kolmogorov distances and negative values are clipped to zero. The black line is the mean position of the wavefront for different system sizes.  We can infer from this that the simulation error arises mainly at the propagating wavefront.}
\label{fig:fisher_kolmogorov}
\end{figure*}

While the means of velocities are instructive in providing general insight into the accuracies of the methods, they do not give complete information about the particle distributions over the entire domain. Thus, for a more accurate analysis, we use the Kolmogorov-Smirnov test \cite{Cao:2006a}, a statistical test used to compare two given distributions. If the particle number distribution for $S_1$\/ within a given subvolume $V_l$\/ at time $t$\/ is $P(X_{l1}(t))$ for a leaping method and $\widetilde{P}(X_{l1}(t))$ for the \NSM{s}, then the Kolmogorov distance between the two distributions is defined as $\mathcal{K}(X_{l1}(t))\!\equiv\!\max|F(X_{l1}(t))\!-\!\widetilde{F}(X_{l1}(t))|$, where $F(x)\!\equiv\!\int_{-\infty}^x P(x)dx$\/ is the cumulative distribution function of $P(x)$.  The reference distribution $\widetilde{P}(X_{l1}(t))$ is also associated with a ``self distance" $\mathcal{S}(X_{l1}(t))$ \cite{Cao:2006a}, which is a measure of the uncertainty associated with building the distribution from a finite set of realizations.  Only if $\mathcal{K}(X_{l1}(t))\!>\!\mathcal{S}(X_{l1}(t))$ can we say that the two distributions are statistically distinct.

In Fig.~\ref{fig:fisher_kolmogorov}, we plot the differences $\mathcal{K}(X_{l1}(5))\!-\!\mathcal{S}(X_{l1}(5))$ over the entire domain $l\!\in\!\{1{\ldots}L\}$ obtained using the full \SPLA{s} and the methods of \MLB{fauthor} and \RBK{fauthor} for various values of $X_{lT}$\/.  A positive value of this difference indicates regions where the solution obtained from the various leaping methods differs, in a statistically significant sense, from the \NSM{s}. These plots reinforce the observations made above: (i) errors arise in \MLB{fauthor} at large populations, (ii) errors arise in \RBK{fauthor} at small populations, and (iii) the full \SPLA{s} is accurate over the entire domain for all system sizes considered.  Moreover, we see that the errors in Figs.~\ref{fig:fisher_kolmogorov}(a)--(c) arise mainly at the propagating wavefront. 

\subsection{Gray-Scott equations} \label{sec:GrayScott}

First studied by Pearson \cite{Pearson:1993}, the Gray-Scott equations
\begin{align}
    \frac{\partial u}{\partial t}& = -uv^2 + F(1-u) + D_u\nabla^2u, \notag \\[6pt]
    \frac{\partial v}{\partial t}& = uv^2 -(F+k)v +D_v\nabla^2v, \label{eq:GrayScottPDE}
\end{align}
describe the spatio-temporal behavior of a two-component reaction-diffusion system.  The equations are of particular interest because they produce a rich variety of spatio-temporal patterns based on the values of $F$\/ and $k$\/.  Here, we set $F\!=\!0.035$, $k\!=\!0.060$, $D_u\!=\!2{\times}10^{-5}$~m$^2$/s and $D_v\!=\!10^{-5}$~m$^2$/s.

\begin{figure}
\centering
\includegraphics[width=3in,height=3in]{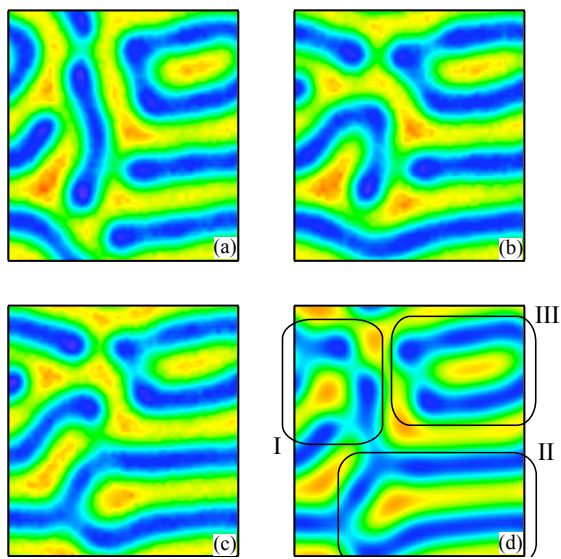}
\caption{Snapshots of the Gray-Scott reaction-diffusion system (\ref{rxndiff:GrayScott}) at $t\!=\!1500$~s obtained using (a) Marquez-Lago, (b) Rossinelli, (c) the full SPLA-SB, and (d) Eqs.~(\ref{eq:GrayScottPDE}). The concentration of $S_1$\/ (plotted above) ranges from 0 (blue) to 1 (red) $M$\/.  The features present in regions I, II and III are compared for different simulation methods. All simulations are performed with the parameters $F\!=\!0.035$, $k\!=\!0.060$,
         $D_u\!=\!2{\times}10^{-5}$~m$^2$/s and $D_v\!=\!10^{-5}$~m$^2$/s.}
\label{fig:gs_plots}
\end{figure} 

We consider a two-dimensional domain of width 0.5~m and length 0.5~m (in say, the y-z plane) and height $H$\/ and divide it into a regular $50{\times}50$ grid ($L\!=\!2500$; $\omega_l\!=\!0.25H$~m$^3$, $\Omega\!=\!N_A\omega_l$) with periodic boundary conditions.  On this domain, we consider the two-component reaction-diffusion system
\begin{alignat}{3}
    S_{l1}& + 2S_{l2} \xrightarrow{k_1} 3S_{l2},        &\quad& l\in\{1{\ldots}L\}, \notag \\ 
    S_{l1}& \xrightleftharpoons[k_{-2}]{k_2} \emptyset, && l\in\{1{\ldots}L\}, \notag \\ 
    S_{l2}& \xrightarrow{k_3} \emptyset,                && l\in\{1{\ldots}L\}, \label{rxndiff:GrayScott} \\ 
    S_{l1}& \xrightleftharpoons[d_1]{d_1} S_{l^\prime 1},   && l\in\{1{\ldots}L\}, &\quad& l^\prime\in\mathbf{C}_l, \notag \\ 
    S_{l2}& \xrightleftharpoons[d_2]{d_2} S_{l^\prime 2},   && l\in\{1{\ldots}L\}, && l^\prime\in\mathbf{C}_l. \notag 
\end{alignat}
If we set the parameters $k_1\!=\!1/\Omega^2$~s$^{-1}$, $k_2\!=\!F$~s$^{-1}$, $k_{-2}\!=\!F\Omega$~s$^{-1}$ and $k_3\!=\!F\!+\!k$~s$^{-1}$, then in the deterministic limit the spatio-temporal evolutions of $S_1$\/ and $S_2$\/ are described by the Gray-Scott equations (\ref{eq:GrayScottPDE}) with $u\!=\!X_1(y,z,t)/\Omega$, $v\!=\!X_2(y,z,t)/\Omega$, $d_1\!=\!D_u/h^2$ and $d_2\!=\!D_v/h^2$.

In the deterministic case, a unique pattern is obtained from Eqs.~(\ref{eq:GrayScottPDE}) for a given set of parameters $\{F,k,D_u,D_v\}$ and initial conditions \cite{Pearson:1993}.  However, the pattern formation behavior can change significantly in the presence of noise \cite{Lesmes:2003}, to the extent that large amounts of internal noise can prevent pattern formation altogether \cite{Wang:2007b}.  We investigate the effects of noise in the Gray-Scott system by performing stochastic simulations using the \SPLA{s}-\SB{s}, \MLB{fauthor} and \RBK{fauthor} methods, but we consciously choose conditions that minimize stochastic effects so that direct comparisons can be made to the deterministic solution.

Initially, we set the concentration of  $S_1$\/ and $S_2$\/ in each subvolume to $1~M$\/ and $0.1~M$ respectively and choose $H$\/ such that $1~M$ corresponds to to 5000 particles. We then apply a perturbation that triggers pattern formation in the reaction diffusion system.  In Fig.~\ref{fig:gs_plots}, we show snapshots of the patterns obtained from the different simulations methods and from the solution of Eqs.~(\ref{eq:GrayScottPDE}) at $t\!=\!1500$~s.

By comparing Figs.~\ref{fig:gs_plots}(a)--(c) to Fig.~\ref{fig:gs_plots}(d), the effects of noise are visually evident.  Rather than the smooth pattern produced in the deterministic case, those obtained from the leaping methods have clear fluctuations.  The effects are small, however, and all of the patterns are superficially similar, although close inspection reveals perceptible differences.  We highlight three regions (I, II and III) in the deterministic solution of Fig.~\ref{fig:gs_plots} to make visual comparison of the patterns easier. The pitchfork-type pattern in region II is present in \SPLA{s} and to some extent in \RBK{fauthor}'s method. That feature is barely recognizable \MLB{fauthor}'s method. Similarly in region I, \SPLA{s}'s pattern is the closer to the deterministic solution than other methods. However, the patterns present in region III are similar in all the simulation methods. From this we argue that the \SPLA{s} pattern in Fig.~\ref{fig:gs_plots}(c) is most similar to the deterministic solution in Fig.~\ref{fig:gs_plots}(d) and that the \MLB{fauthor} pattern in Fig.~\ref{fig:gs_plots}(a) is most dissimilar.  Since we consciously aimed to minimize the effects of stochasticity in the pattern formation, these results imply that the most faithful description of the system dynamics is given by the \SPLA{s}.

In order to ascertain why this is, we compare the time steps taken by the \SPLA{s} to those for the \MLB{fauthor} and \RBK{fauthor} methods. The main result (plot not shown) is that the full \SPLA{s} generally takes smaller time steps than the other methods, explaining why it gives more accurate results.

It is important to note that our analysis of the Gray-Scott system is limited due to the large number of total events in the system.  With 2500 subvolumes, each with four nearest neighbors, there are a total of $10^4$ reactions and $2{\times}10^4$ diffusion events that must be taken into account. As such, a single \SPLA{s} simulation of 1500~s took $1.43$~h to complete.  \MLB{fauthor} and \RBK{fauthor} simulations took a comparable amount of time ($0.36$~h and, $0.92$~h respectively). This is an important result because it exemplifies a serious shortcoming of the spatial $\tau$\/-leaping approach in general.  Although leaping is beneficial in allowing multiple event firings at each simulation step, the high cost of $\tau$\/~selection severely limits the applicability of the approach in the face of large event numbers, as is common in spatially-discretized systems.  Thus, in order to make the approach practicable, improving the efficiency of the method is of paramount importance.  We discuss this issue in more detail in Sec.~\ref{sec:discussion}. 

\section{Discussion and Conclusion} \label{sec:discussion}

We have presented the spatial partitioned-leaping algorithm as an accurate formulation of the leaping approach for reaction-diffusion systems on discretized grids. Our primary contributions have been to correctly enumerate all of the events that must be considered during the time-step calculation process and to recast the \RB{l} and \SB{l} $\tau$\/-selection formulas \cite{Cao06:newStep, Harris06} within a spatial context (Tables~\ref{table:RBTauSelec} and \ref{table:SBTauSelec}). The main differences between these formulas and those used in prior implementations of spatial $\tau$\/-leaping \cite{Lago07, Rossi08} are that reaction and diffusion events are considered together and the effects of incoming diffusion are properly taken into account.  Both aspects are crucial for an accurate spatial leaping implementation and we have shown, through numerical examples, how improper consideration can lead to the introduction of error or a reduction in efficiency, depending on the specifics of the system being studied. We have also shown the implications, in terms of accuracy, of not providing a mechanism for transitioning to a \ES{l} method in the limit of small populations.

Furthermore, we have shown that the \SB{l} $\tau$\/-selection procedure, besides being inherently less costly per calculation than the \RB{l} procedure (because of the lack of rate derivatives \cite{Cao06:newStep}), will generally require far fewer total calculations for spatial systems than the \RB{l} approach. This is because the total number of events in a discretized system will often far exceed the total number of species.  Species-based $\tau$\/-selection will thus be the preferred choice in most situations. Exceptions include cases where rate constants are time dependent, e.g., if environmental quantities such as temperature and volume vary in time (\SB{l} $\tau$\/-selection assumes time-invariant rate constants).  In such situations, modified forms of the \RB{l} $\tau$\/-selection formulas will be required.

Inclusion of the \ES{l} classification in the algorithm brings along problems of classification cascade, induced by events at the edge of diffusing fronts, which eventually results in an unnecessarily small time step and, hence, a significant reduction in efficiency. This phenomenon has been observed previously for a well-mixed biochemical system involving binding of transcription factors to individual genes \cite{Harris09} and is a shortcoming of the PLA in general.  Here, we have attenuated this problem to an extent by restricting the classification of diffusion events as \ES{l} when the population of the diffusing species exceeds a pre-specified threshold (i.e., 100). This approach is \textit{ad~hoc\/}, however, and we have demonstrated the need to develop a more general approach that can handle all cases. Work is currently underway in this direction.

A shortcoming of the \SPLA{s}, and other spatial leaping methods in general, is the strong dependence of the computational cost on the total number of events in the system. This is a well-known problem for stochastic simulation algorithms \cite{Petti05} and is exemplified by the large amount of time taken to analyze the moderately complex Gray-Scott system (involving $3\!\times\!10^4$ unique events involving 5000 unique species). These methods remain constrained by the fact that one $\tau$\/-selection calculation must ultimately be performed for each event (\RB{l}) or species (\SB{l}) present in the system. In order to make the approach practicable, a solution to this problem is clearly required.  A computational approach can be to parallelize the algorithm, parsing out the computational effort across multiple machines.  Many aspects of the SPLA are indeed parallel in nature, such as $\tau$\/~selection, event classification and event update.  From an algorithmic perspective, Anderson's post-leap checking procedure \cite{Ander08} may provide some relief in that it obviates the need to perform the expensive pre-leap calculations. Pettigrew and Resat \cite{Petti07} have proposed an approximate post-leap checking procedure that might prove useful as well.

Another possibility is to fundamentally reduce the number of $\tau$\/~selection calculations by performing them on \emph{groups\/} of events rather than on individual events or species. The challenge, however, is that in contrast to the exact-stochastic case (Sec.~\ref{sec:inhomogeneous}), it is not permissible within the context of a leaping algorithm to group arbitrary sets of events and then perform $\tau$\/~selection on the group.  This is because the leap condition Eq.~(\ref{eq:leapCondition}) applies at the level of individual events, not groups.  Basically, there is no guarantee that a given change in the summed propensity of the group will translate into equivalent changes in the propensities of the events that comprise the group.  However, it may be possible to identify special types of groups in which this is, in fact, the case.  This type of grouping, based on event type rather than on location, is fundamentally different from that used in typical spatial simulation methods.  It also differs from the type of grouping used in the multinomial $\tau$\/-leaping method of Pettigrew and Resat \cite{Petti07}, a well-known binomial $\tau$\/-leaping variant.  We are actively pursuing this avenue of research.

Compounding the problem of \ES{l} event classifications is that \SPLA{s} transitions to \NRM{s}, which is an inefficient \ES{l} method for spatial simulations. Ideally, the method would segue to an efficient spatial SSA formulation such as the NSM.  However, the NSM, which is based on grouping events by subvolume, does not fit naturally into the framework of the \SPLA{s} for the reasons cited above, i.e., $\tau$\/~selection cannot be applied at the level of groups. \MLB{fauthor} incorporate the \NSM{s} into their spatial $\tau$\/-leaping method by classifying subvolumes as \ES{l} if $a_{l0}(t)\!\lesssim\!10$ (Sec.~\ref{sec:variants}), emulating the approach taken by Gillespie, Petzold and co-workers \cite{Gillesp01, Gillesp03, Cao06:newStep}.  We could employ a similar approach in the \SPLA{s}. However, it is our hope that a more natural method of transition will arise from our attempts to incorporate grouping generally into the leaping methodology.

Our development of the \SPLA{s} is significant in that it represents a ``gold standard" in terms of accuracy against which future enhancements and extensions to the spatial $\tau$\/-leaping approach can be compared. As a straightforward implementation of spatial leaping, the method is not maximally optimized in terms of efficiency nor is it meant to be.  However, it does achieve the maximum possible gains in efficiency for a method that accurately employs pre-leap $\tau$\/~selection at the level of individual events by considering only those events that are of consequence to the calculation.  These include local reactions and outgoing and incoming diffusion events to and from neighboring subvolumes. We hope that future innovations addressing the challenges highlighted here will help to further improve the leaping methodology and make stochastic simulations of complex systems practicable.

\section{Acknowledgments} \label{sec:acknowledgments} 
 
We thank the anonymous reviewers for their useful inputs and the Semiconductor Research Corporation for financial support. 



\printfigures
\printtables
\end{document}